\documentstyle[12pt]{article}

\catcode`\@=11

\def\addcontentsline#1#2#3{\relax}

\def\fnum@figure{Figure \thefigure}
\def\fnum@table{Table \thetable}
\newcounter{figcaption}
\def\thefigcaption{\arabic{figcaption}}
\def\fnum@figcaption{{\bf Fig. \thefigcaption.}}
\def\figcaption{%
{\parindent 0pt \bf Figure Captions} \par \vskip 10pt
\list{\fnum@figcaption}
{\leftmargin 5em \labelwidth\leftmargin\advance\labelwidth-\labelsep
\def\makelabel##1{##1\hfil} \usecounter{figcaption}}%
}

\newcommand{\rmd}{{\rm d}}

\newcommand{\rmi}{{\rm i}}

\renewcommand{\Im}{{\cal I}\!{\sl m}\,}

\newcommand{\Ham}{{\cal H}}

\newcommand{\dags}{^\dagger}

\def\braket#1{\left\langle#1\right\rangle}

\def\brakets#1{\langle#1\rangle}

\def\simle{\mathrel{\mathpalette\@versim<}}   
\def\simge{\mathrel{\mathpalette\@versim>}}   
\def\@versim#1#2{\lower2.5pt\vbox{\baselineskip0pt \lineskip-.5pt
   \ialign{$\m@th#1\hfil##\hfil$\crcr#2\crcr\sim\crcr}}}

\newcommand{\bequ}{ \begin{equation} }
\newcommand{\eequ}{ \end{equation} }
\newcommand{\barr}{ \begin{array} }
\newcommand{\earr}{ \end{array} }
\newcommand{\beqarr}{ \begin{eqnarray} }
\newcommand{\eeqarr}{ \end{eqnarray} }

\newcommand{\baralpha}{ \begin{eqnal} \beqarr}
\newcommand{\earalpha}{ \eeqarr \end{eqnal}}

\catcode`\@=12

\def\MFreq{\rmi \omega_n}

\def\vecn{\vec m}

\def\LSMO{La$_{1-x}$Sr$_x$MnO$_3$}

\begin{document}

\begin{Large}
\bf
\noindent
Magnetic and Transport Properties of \\
the Kondo Lattice Model \\
 with Ferromagnetic Exchange Coupling%
\footnote{%
to be published in Proc. 17th Taniguchi International Conference
 on "Spectroscopy of Mott Insulators and Correlated Metals", edited by
A. Fujimori and Y. Tokura (Springer Verlag)}

\end{Large}

\vspace{3.1mm}
\noindent
N.  Furukawa

\vspace{2.6mm}
{\noindent
  Institute for Solid State Physics, University of Tokyo,\\
 Roppongi 7-22-1, Minato-ku, Tokyo 106, Japan
}

\vspace{11.0mm}
\noindent
{\bf Abstract.} The Kondo lattice model with Hund's ferromagnetic spin
coupling is investigated
as a microscopic model of the perovskite-type $3d$ manganese oxide
$R_{1-x}$$A_x$MnO$_3$ where $R$ and $A$ are rare earth element and
alkaline earth element, respectively.
We take the classical spin limit
 $S=\infty$ for the simplicity of the
calculation, since the quantum exchange process seems to be irrelevant
in the high temperature paramagnetic phase.
Magnetic and transport properties of the system are calculated.
In the hole doped systems, ferromagnetic instabilities are observed as the
temperature is lowered.
The giant magnetoresistance of this model
is in  excellent agreement with the experimental data of {\LSMO}.

\vspace{-1.1mm}

\section{Introduction}
\hspace{\parindent}
Strong correlation in electron systems is one of the
unsolved problems in the field of the condensed matter physics
which has been studied over several decades.
Especially, after the discovery of the high-$T_{\rm c}$ materials,
many theoretical and experimental works have been performed
to approach this problem.

One of the groups of materials that the strong correlation effect
is observed in its physical properties is the
 $3d$ transition-metal oxides with perovskite-like lattice structures
$(R,A)M$O$_3$.
In these materials,  one can control the
number of $3d$ electrons by changing the species of the transition-metal $M$.
By choosing the component of $R$ and $A$,
rare earth element and alkaline earth element, respectively,
we can  change electron hopping energy and carrier concentration.
One of the novel feature in this group of materials is that
the carrier doping is possible without destroying the
fundamental network of $M$O$_3$, so that the effect of randomness
can be minimized. Prominent spin-charge coupled properties
in these materials near the metal-insulator transition
due to electron correlation are observed.

As an example, the effective mass of the carrier doped
LaTiO$_3$ shows anomalous enhancement \cite{Tokura93},
which is considered to be the effect of the strong correlation
in the vicinity of the Mott transition.
{}From the theoretical point of view,
the Hubbard model near half-filling is examined as a simplified model for
the doped LaTiO$_3$. The singularity
in the charge mass is observed at
$n\to 1$ \cite{Furukawa93,Rozenberg94}.

Recently,
transport properties for filling-controlled
single crystals of manganese oxides $(R,A)$MnO$_3$
have been investigated systematically
\cite{Tokura9x,Tomioka9x}.
In this materials, $3d$ electrons are considered to form
both localized spins and itinerant electrons which are
coupled  each other by Hund's interaction.
Several complex phenomena due to strongly coupled
spin, charge and lattice degrees of freedom are observed.

For  moderately
doped samples of {\LSMO}, there exists a ferromagnetic metal state
below $T_{\rm c}$
due to double exchange mechanism 
\cite{Zener51,Anderson55,deGennes60},
while at $x\sim 0$, the antiferromagnetic insulator phase is observed.
At moderately doped region  \cite{Tokura9x},
a sharp drop is
observed in the resistivity below the magnetic transition
temperature $T_{\rm c}$.
It is also reported that the field-induced magnetic moment
also increases the conductivity. Namely, the giant magnetoresistance (GMR)
with negative sign is observed in this material.
In (Pr,Ca)MnO$_3$, the phase transition from a spin glass insulator to a
ferromagnetic metal is observed by inducing the magnetic field
at low temperatures. Under certain conditions, the
change of resistivity is larger than ten orders of magnitude
when the external magnetic field is applied \cite{Tomioka9x}.
The interesting points of these phenomena are that
they show a new type of magnetotransport feature as well as
their possibilities to be applied to devices.

In this paper, we study thermodynamical properties of
a microscopic model which effectively describes
the nature of $(R,A)$MnO$_3$.
One of our aims is to calculate the conductivity
and the magnetoresistance
in order to compare with the experimental data in {\LSMO}.

\section{Model}
\hspace{\parindent}
In a Mn$^{3+}$ ion, $t_{\rm 2g}$ orbitals and $e_{\rm g}$ orbital
are considered to be occupied with three and one $3d$ electrons,
respectively. The doped holes enter the band constructed from
strongly hybridized Mn $e_{\rm g}$ and O $2p$ orbitals.
We consider that the $t_{\rm 2g}$ electrons are nearly localized
and mutually coupled ferromagnetically due to Hund's rule, so that they form
$S=3/2$ Heisenberg spins.
They are also coupled with $e_{\rm g}$ electrons by Hund's coupling.
Then, as a model Hamiltonian of this system, the Kondo lattice model (KLM)
in three dimensions,
\bequ
 \Ham = - \sum_{ij,\sigma} t_{ij}
        \left( c_{i\sigma}\dags c_{j\sigma} + h.c. \right)
    -J \sum_i \vec \sigma_i \cdot \vec S_i
\eequ
with spin $S=3/2$ and  the Hund's ferromagnetic
 coupling $J>0$ has been proposed \cite{Kubo72}.
Here, the Pauli matrices $\vec \sigma_i
 = (\sigma_i^x,\sigma_i^y,\sigma_i^z)$
represent the spin of itinerant electrons while $\vec S_i$ denotes
the localized spin.
The Hund's coupling parameter $J$ is estimated to be
larger than the band width $W$ of the itinerant electrons,
so that the system is in the strong coupling regime.
In order to make a clear distinction from the
antiferromagnetically coupled Kondo lattice model,
we sometimes refer to this model as the $e$-$t$ model \cite{SawatzkyPC},
since it is constructed from $e_{\rm g}$ and $t_{\rm 2g}$ electrons.
The carrier number of the band electron for
$R_{1-x}A_x$MnO$_3$ is considered to be $n\simeq 1-x$.

Since we consider the case where
 the localized spin is in a high-spin state with the ferromagnetic coupling,
the effect of quantum exchange
seems to be unimportant in the paramagnetic phase
where the thermal fluctuation of spins is dominant.
Thus we consider the infinite high-spin limit $S=\infty$
so that the localized spins are classical rotators.
The Hamiltonian is described as
\bequ
  \Ham =
  - \sum_{ij,\sigma} t_{ij}
        \left(  c_{i\sigma}\dags c_{j\sigma} + h.c. \right)
    -J \sum_i \vec \sigma_i \cdot \vec m_i,
    \label{HamSinfty}
\eequ
where $ \vec m_i = (m_i{}^x, m_i{}^y, m_i{}^z)$ and $|\vec m|^2 = 1$.
The partition function is given by
\bequ
  Z = {\rm Tr}_{\rm S} {\rm Tr}_{\rm f}
      \exp[ -\beta(\Ham-\mu \hat N)].
	\label{defPartFun}
\eequ
Although the Hamiltonian (\ref{HamSinfty})
is only constructed from one-body terms for the fermion degrees of freedom,
thermodynamic
properties of this system
are still not easy to obtain
since we have to take the trace over all the localized spins
in eq.~(\ref{defPartFun}).
One of the methods to calculate the thermodynamical quantities is
the Monte Carlo method for  cluster systems combined with
the finite size scaling analysis.
Our numerical results will be shown elsewhere.
In this paper, we give another approach to obtain the
thermodynamic limit of the system.

In order to calculate the
thermodynamic properties in a controlled manner,
we take
the limit of infinite dimension $D=\infty$ \cite{Metzner89}.
This corresponds to the limit of large coordination number where we
may neglect  site off-diagonal terms in  self-energies
and vertex corrections.
In this limit, the problem is
reduced to the single-site problem coupled with a
 dynamical mean field \cite{Georges92}.
In ref.~\cite{Furukawa94}, the present author have
 shown that the one-body Green's function
of the $D=\infty$ and $S=\infty$ KLM can be
calculated exactly.
The Green's function is calculated exactly as
\bequ
  \tilde G(\MFreq) =
  \braket{ \left(  \tilde G_0^{-1}(\MFreq) +
            J \vecn \vec \sigma \right)^{-1}}_{\vecn},
\eequ
where $\braket{\cdots}_{\vecn}$ represents the thermal average over
$\vecn$.
We determine the Weiss field $\tilde G_0$
self-consistently.

In the paramagnetic phase,
we have
\bequ
  \tilde G(\MFreq) =
    \braket{
      \frac{
            \tilde G_0(\MFreq)^{-1}  - J \vecn \vec \sigma
      }
      {  \tilde G_0(\MFreq)^{-2} - J^2 |\vecn|^2 }
    }_{\vecn}
     = \frac{  \tilde G_0(\MFreq)^{-1} }
             {  \tilde G_0(\MFreq)^{-2} - J^2},
   \label{GFunAve}     \label{GFunSolution}
\eequ
since $\braket{\vecn} = \vec 0$ in the paramagnetic phase.
The self-energy is then given by
\bequ
  \Sigma(\MFreq) =
    \tilde G_0^{-1}(\MFreq) - \tilde G^{-1}(\MFreq)
   = J^2 \tilde G_0(\MFreq).
    \label{SelfEnergy}
\eequ
{}From the derivation, it is clear that, as long as we
restrict ourselves to the paramagnetic phase, the Green's function
of the present model is the same as that of
the system with the Ising spin
$\vecn = (0,0,\pm1)$. Therefore, the paramagnetic solution has
the close relationship between that of the Falicov-Kimball model
in $D=\infty$.

Using the Kubo formula, the optical conductivity is
calculated from the equation \cite{Moller92,Pruschke93a}
\beqarr
  \sigma(\omega) &=& \sigma_0
  \sum_\sigma \int \rmd \omega' \int \rmd \epsilon \
     W^2 N_0(\epsilon) \nonumber \\
  & & \quad \times \
      A_\sigma(\epsilon,\omega') A_\sigma(\epsilon,\omega'+\omega)
   \frac{ f(\omega') - f(\omega'+\omega)}{\omega},
   \label{Optcond}
\eeqarr
where
\bequ
 A_\sigma(\epsilon,\omega')
 = - \frac{1}{\pi} \Im G_{\sigma}(\epsilon,\omega'+ \rmi\eta)
\eequ
is the spectral function and $f(\omega)$ is the Fermi distribution
function.
Here we used the fact that the vertex corrections vanish in the limit of
infinite dimension.
The constant $\sigma_0\sim ({ e^2 a^2}/{\hbar}) \cdot ({N}/{V})$
gives the unit of the conductivity
where $a$ is the lattice constant.

\section{Results at $D=\infty$}
\hspace{\parindent}
Calculations are performed using the Lorentzian density of states
as well as the semi-circular density of states
with the bandwidth $W\equiv 1$. 
We express the induced magnetization by $M = \brakets{m_i{}^z}$.

It has been shown that the analytical structure of the Green's function
of the KLM in $D=\infty$ and $S=\infty$ is the same as that of
the Falicov-Kimball model in $D=\infty$ which
has been studied intensively \cite{Furukawa94}.
As a typical example,
the imaginary part of the self-energy at the fermi level is finite
$  \Im \Sigma(0) \ne 0$ in the paramagnetic phase \cite{Moller92,Si92}.
As the magnetic moment is induced,
the thermal fluctuation of spins decreases
which causes the decrease
 of the imaginary part of the
self-energy.

In Fig.~\ref{FigDOS-M}, we show the density of states
of the up-spin particle at $J=4$ under the
 magnetic field for various values of the induced magnetization $M$.
At $M=0$, the density of states splits into lower and upper band
in the case $J \gg W$, which has the similar behavior
as seen in the Hubbard approximation.
For the semi-circular density of states, the gap opens at
$ J_{\rm c} = 0.5 W$  \cite{vanDongen92}.
Therefore,
 we have the metal-insulator transition at half-filling
in the strong coupling region.
The Kondo resonance peak is not observed in this model.
As the local spin is polarized, the weight at lower band increases
for up-spin fermion.

In Fig.~\ref{FigOptCond}, we show the optical conductivity $\sigma(\omega)$
at $n=0.8$
for various values of $M$, which is calculated from eq.~(\ref{Optcond}).
At paramagnetic phase $M=0$, we see a peak structure at $\omega \sim 2J$
as well as the Drude part at $\omega \sim 0$.
The peak at $\omega \sim 2J$ corresponds to the excitation process
from the lower band to the upper band of the split density of states.
As the magnetization is induced, the peak at $\omega \sim 2J$ diminishes
and the Drude part develops. This is explained by the change of
the shape in the density of states from the two-peak structure at $M=0$ to
the single-peak structure at $M=1$.
{}From the experimental point of view,
the optical conductivity in moderately doped
{\LSMO} shows the similar behavior when the temperature-induced
magnetic moment develops \cite{TokuraPC}.

In Fig.~\ref{FigRhoRho0}, we show the resistivity $\rho$ scaled by its
zero-field value $\rho_0$ as a function of $M$.
This curve
reproduces the experimental data of GMR  in {\LSMO} \cite{Tokura9x}
excellently.
At $M \ll 1$ we see
\bequ
  \frac{\rho(M)}{\rho(M=0)} = 1 - CM^2,
\eequ
where $C \sim 4$ at $J=4$ and $n=0.825$.
The value of $C$ decreases as the hole is doped \cite{Furukawa94},
which is in qualitative agreement with the experimental data \cite{TokuraPC}.

\section{Discussion}
\hspace{\parindent}
The calculation of the present model
in the strong coupling limit $J \gg W$ shows that
the quasi-particle excitation is incoherent in the paramagnetic phase
since $\Im\Sigma \sim J^2/W$ is satisfied.
The above results imply that the system has a short coherence length.
Therefore, the effective single-site treatment of the $D=\infty$ system
is justified in the strong coupling limit.
The time scale of the quantum spin-flip process
is estimated by $1/T_{\rm K} \sim 1/W$  at $J \gg W$,
which is relatively longer than the
quasi-particle lifetime $1/\Im\Sigma \sim W/J^2$.
Then, it is also justified
to take the limit of  $S=\infty$.

Thus,
the GMR in {\LSMO} is quantitatively reproduced by the present calculation.
In order to understand the glassy behavior in (Pr,Ca)MnO$_3$,
where the band width of $e_{\rm g}$ electrons are relatively narrower,
it seems to be necessary to introduce antiferromagnetic
exchange coupling between localized spins and impurity potentials
which has been neglected in {\LSMO},
so that the frustration effect takes place.
Such microscopic analysis of (Pr,Ca)MnO$_3$ and other related materials
will be reported elsewhere.

\section*{Acknowledgments}
\hspace{\parindent}
The author would like to thank Y. Tokura, T. Arima, F.~F. Assaad
and M. Imada for fruitful discussions and comments.

\begin{figcaption}

\item Density of states of up-spin particle
at $J=4$ under magnetic field.
\label{FigDOS-M}

\item Optical conductivity at $J=4$ $n=0.8$ for various values of $M$.
\label{FigOptCond}

\item Magnetization dependence of $\rho/\rho_0$ at $\braket{n}=0.8$.
\label{FigRhoRho0}

\end{figcaption}

\begin{thebibliography}{10}

\bibitem{Tokura93}
Y. Tokura, Y. Taguchi, Y. Okada, Y. Fujishima, T. Arima, K. Kumagai and Y. Iye,
  Phys. Rev. Lett. {\bf 70}, 2126 (1993).

\bibitem{Furukawa93}
N. Furukawa and M. Imada, J. Phys. Soc. Jpn. {\bf 62}, 2557 (1993).

\bibitem{Rozenberg94}
M.~J. Rozenberg, G. Kotliar and X.~Y. Zhang, Phys. Rev. {\bf B49},
10181 (1994).

\bibitem{Tokura9x}
Y. Tokura, A. Urushibara, Y. Moritomo, T. Arima, A. Asamitsu, G. Kido and N.
  Furukawa: J. Phys. Soc. Jpn. {\bf 63} (1994) 3931.

\bibitem{Tomioka9x}
Y. Tomioka, A. Asamitsu, Y. Morimoto and Y. Tokura, preprint.

\bibitem{Zener51}
C. Zener, Phys. Rev. {\bf 82}, 403 (1951).

\bibitem{Anderson55}
P.~W. Anderson and H. Hasegawa, Phys. Rev. {\bf 100}, 675 (1955).

\bibitem{deGennes60}
P.~G. de~Gennes, Phys. Rev. {\bf 118}, 141 (1960).

\bibitem{Kubo72}
K. Kubo and N. Ohata, J. Phys. Soc. Jpn. {\bf 33}, 21 (1972).

\bibitem{SawatzkyPC}
G.~A. Sawatzky, private communication.

\bibitem{Metzner89}
W. Metzner and D. Vollhardt, Phys. Rev. Lett. {\bf 62}, 324 (1989).

\bibitem{Georges92}
A. Georges and G. Kotliar, Phys. Rev. {\bf B45}, 6479 (1992).

\bibitem{Furukawa94}
N. Furukawa, J. Phys. Soc. Jpn. {\bf 63}, 3214 (1994).

\bibitem{Moller92}
G. {M\"oller}, A.~E. Ruckenstein and S. {Schmitt-Rink}, Phys. Rev. {\bf B46},
7427  (1992).

\bibitem{Pruschke93a}
T. Pruschke, D.~L. Cox and M. Jarrel, Phys. Rev. {\bf B47}, 3553 (1993).

\bibitem{Si92}
Q. Si, G. Kotliar and A. Georges, Phys. Rev. {\bf B46}, 1261 (1992).

\bibitem{vanDongen92}
P.~G.~J. van Dongen, Phys. Rev. {\bf B45}, 2267 (1992).

\bibitem{TokuraPC}
A. Urushibara, Y. Morimoto, T. Arima, A. Asamitsu, G. Kido and Y. Tokura,
  preprint.

\end{thebibliography}
\end{document}